# Deterministic End-to-End Transmission to Optimize the Network Efficiency and Quality of Service: A Paradigm Shift in 6G

Xiaoyun Wang, Shuangfeng Han, Zhiming Liu, and Qixing Wang
China Mobile Research Institute

**Abstract:** Toward end-to-end mobile service provision with optimized network efficiency and quality of service, tremendous efforts have been devoted in upgrading mobile applications, transport and internet networks, and wireless communication networks for many years. However, the inherent loose coordination between different layers in the end-to-end communication networks leads to unreliable data transmission with uncontrollable packet delay and packet error rate, and a terrible waste of network resources incurred for data re-transmission. In an attempt to shed some lights on how to tackle these challenges, design methodologies and some solutions for deterministic end-to-end transmission for 6G and beyond are presented, which will bring a paradigm shift to the wireless communication networks.

*Keywords*— 6G, application layer, transport layer, quality of service, deterministic transmission,

## I. Efforts toward better End-to-end Wireless Service Provision

*End-to-end wireless service provision process:* In the wireless communication networks, the provision of mobile service requires data processing in different layers of the Open System Interconnection (OSI) reference model. The data packet of diversified mobile services will first be source-coded at the application layer and transmitted via transport layer protocols like Transmission Control Protocol (TCP) or User Datagram Protocol. Then the data packet goes through the Internet Protocol (IP) network layer and arrives at the core network (CN), where the quality of service (QoS) [1] parameters will be configured to ensure the mobile services will be provisioned with sufficiently satisfactory quality. The CN estimates the QoS requirements of each data flow, decides on whether the data flow type is guaranteed bit rate (GBR), delay critical GBR, or non-GBR, and estimates the packet delay budget, packet error rate, and so on. The radio access network (RAN) will continue to schedule the transmission of these data packets via the air interface according to the QoS requirements.

*Efforts toward better wireless communication networks:* To satisfy the ever increasing requirements of mobile services, the wireless communication industry has developed from the first generation analog communication networks to the current fifth generation (5G) networks [2,3]. Now, the global attention has been shifting from 5G to the future 5.5G and 6G. Toward 5.5G, the standard body like 3GPP started comprehensive studies in release 18 to adopt more new features, including full duplex, wireless artificial intelligence and machine learning, network energy saving, XR service enhancement [4], etc. Release 19 may well continue the studies for 5.5G. Extensive studies have also been conducted in both the industry and academia on the 6G scenarios, use cases, and key technologies [5-7]. For example, several research directions for 6G have been identified, including Terahertz communications, reflective intelligent surface, artificial intelligence and deep learning driven network architecture and transmission technology in physical and higher layer. In the recent ITU-R recommendation [8], the framework and overall objectives of the future development of IMT for 2030 and beyond has been outlined. IMT-2030 is expected to support enriched and immersive experience, enhanced ubiquitous coverage, and enable new forms of collaboration.

*Efforts toward better application layer and TCP/IP layer:* Toward a better end-to-end performance, other parts in the OSI model have also been evolving continuously.

1) *Application layer optimization:* Take the video coding and decoding for example. Video codecs for multimedia services have been upgraded for many years. For example, the Moving Picture Expert Group (MPEG) immersive video standard [9], the latest addition to the MPEG-I suite of standards, is designed to support virtual and extended reality applications that require six degrees of freedom visual interaction with the rendered scene. The MPEG-5 low complexity enhancement video coding [10] works in combination with other codecs, to produce a more efficiently compressed video. In recent years, artificial intelligence has been adopted in the codec design to more accurately capture the key message in the information source to further increase the compression ratio.

2) *Transport layer optimization:* In the transport layer, applications using TCP in the wireless or wired communication are often bottle-necked by the handshake mechanism, which may undesirably incur delay, particularly in the time-sensitive communication scenario (e.g. streaming live video). Many congestion control algorithms have been developed. For example, the TCP Westwood [11] uses packet loss rate as a key indicator of network congestion to adjust the TCP transmission window, and is particularly useful when unexpected losses due to radio channels are misinterpreted as congestion, resulting in window reduction. Artificial intelligence/machine learning technologies have recently been introduced to enhance the TCP performance [12,13].

3) *Internet protocol layer optimization:* Traditional IP protocols support reliable service data delivery, without providing strict QoS guarantees. To frame the development of new technologies for deterministic IP networks, the Internet Engineering Task Force (IETF) and Institute of Electrical and Electronics Engineers (IEEE) Time Sensitive Network (TSN) have specified target requirements and potential solutions for guaranteed bandwidth, bounded end-to-end latency and bounded jitter [14].

## II. Fundamental Challenges to the Wireless Communication Networks

Despite all the above efforts toward better end-to-end service performance, there are still several fundamental limitations of the current wireless communication networks. As shown in Fig.1, the traditional QoS management is confined to the wireless networks (i.e. the part below transport/IP layer, as specified in [15]). There is a clear lack of efficient QoS management in the application layer, transport layer and IP layer, which has been and will be hampering end-to-end mobile service optimization.

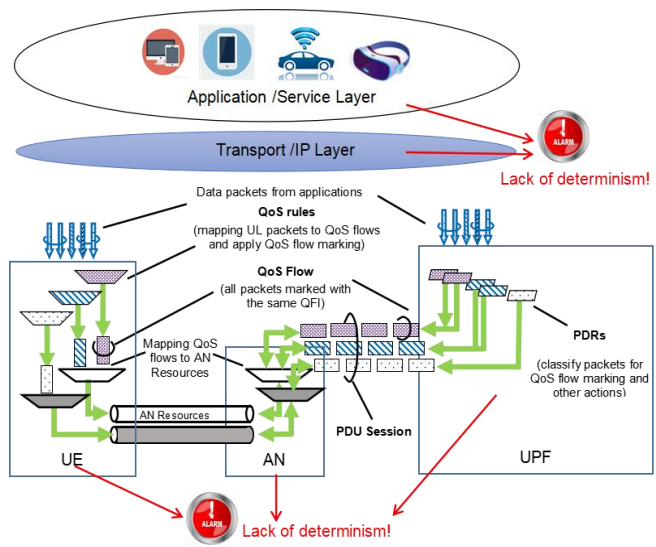

Fig. 1. Limitations of the end-to-end wireless networks

*Lack of determinism in the application layer:* Significant progress has been made in the mobile applications to better encode and decode the source information, e.g. to achieve higher compression ratio while maintaining satisfactory quality. However, dynamic variations in space, time, frequency domains are often observed in the wireless channels, especially in the high mobility scenarios. These dynamic variations often lead to re-transmission in media access control (MAC) layer, radio link control (RLC) and transport layer. Currently, the applications don't have sufficient access to the TCP and IP performance, the wireless channels and QoS guarantee capability at the RAN side. This means the data format of the application layer, if not configured properly, may not be satisfactorily supported by the subsequent layers.

*Lack of determinism in the transport layer:* Wireless TCP has become the bottle of the wireless communications, mainly due to the following reasons.

1) *Difficulty in identification of the conjestion reasons:* The traditional TCP schemes work well for traditional Internet traffic since packet loss is mainly caused by congestion. However, in wireless communication networks, the packet loss is most likely caused by variations of wireless channels, rather than true congestion. The erroneously triggered congestion control can result in lower bandwidth utilization.

2) *Difficulty in adjustment of the transmit data packet size:* TCP protocols use the probing mechanism to change the packets transmission size with inherent blindness towards the buffer status at the RAN, which is determined by the current buffer status and the scheduled transmission data size during the next TCP window. This is because the TCP layer does not know the scheduling algorithms at the RAN and has no clue how many data will be transmitted for each mobile user, thus being unable to adjust the packet transmission speed accordingly.

3) *Difficulty in QoS guarantee in TCP layer:* The transport layer manages data flows uniformly and maintains best-effort fairness for all the applications. Consequently, the TCP layer is faced with fundamental challenges in satisfying various QoS requirements like high data-rate, constant bit-rate, delay-tolerant, or high reliability. It is obvious that poor or no QoS guarantee at the TCP layer will inevitably degrade the end-to-end service quality.

*Lack of determinism in the wireless QoS management:* For information security and privacy purpose, the mobile over-the-tops would not allow more exposure of the application data to the operators. Consequently, it is challenging for the operators to obtain services characteristics and to make a proper prediction and guarantee of QoS with optimized resource allocation, thus leading to a degradation of network efficiency. To make it worse, even if the CN identifies QoS requirements accurately, the configured QoS parameters may not be achievable in the RAN. This is because the temporal and geographical distribution of the mobile services may fluctuate, and the wireless channels are dynamically changing almost all the time in the mobile communications. No matter how smart and powerful the resource scheduler at the RAN is, there is a considerable possibility that the mobile service may not be supported due to lack of radio resources. This is the essential reason for lack of determinism in the wireless QoS management.

*Lack of determinism in the IP layer:* There exist several challenges to providing end-to-end latency and bounded jitter guarantees in the IP layer. Firstly, due to the multiplexing nature of the data bursts, the varying elasticity of mobile traffic needs to be controlled at the ingress, thus incurring an extra latency depending on the burst size. Secondly, generally a very large number of devices and users will use the IP networks simultaneously. It would be difficult to manage the delay and jitter of each mobile traffic after being routed many times.

Based on the above analysis, due to the inherent loose coordination between different layers in the OSI service model, the end-to-end transmission is inevitably uncertain and unpredictable, thus rendering the end-to-end QoS guarantee unreliable and challenging. This also results in very low network efficiency. Unfortunately, this issue has not received due attention for various reasons in the wireless communication industry.

## III. NEW PARADIGM OF WIRELESS COMMUNICATIONS WITH DETERMINISTIC END-TO-END TRANSMISSION

Deterministic network is especially important for the emerging time-critical applications, such as industrial automation, smart grids, and telesurgery. In order to tackle the fundamental challenges faced with the wireless communication networks to achieve maximized network efficiency and quality of service, it would be highly motivated to introduce certain determinism into the end-to-end data transmission. The following design methodologies for deterministic end-to-end transmission need to be taken into considerations in the design of 6G and beyond.

*Deterministic QoS jointly determined for each layer:* In the current 5G networks, the QoS parameters are determined in the CN, which has no idea whether these parameters could by supported in the RAN, especially when the mobile resources are limited in the peak wireless traffic hours. Therefore, coordination between the CN and RAN is necessitated, and the final QoS parameters (determined in the CN, RAN or other management platforms) should be sufficiently supported by the RAN scheduling algorithms. For a better end-to-end QoS guarantee, the QoS parameters for each layer need to be jointly determined. One design example is given to illustrate how to jointly determine the QoS parameters:

1) The CN or other management entity in the network estimates the QoS requirements of each data flow. Traditionally, these QoS requirements are service targets of the wireless networks. For the future design, the QoS requirements should be for the end-to-end communication.

2) The CN or other management entity in the network will coordinate the end-to-end QoS requirements for all the layers, including the application layer, transport layer, IP layer, CN, RAN and the user equipment (UE). For example, the delay budget will be partitioned into different layers to jointly fulfill the end-to-end delay requirements.

3) The RAN calculates whether the wireless QoS requirements of all users will be supported in RAN. If yes, the QoS parameters will be used at the RAN as the scheduling targets. If not all the users' QoS requirements will be supported, even with the smartest scheduling algorithm, the RAN will suggest QoS degradation to CN based on some fairness metric, e.g. the one that ensures same ratio of scheduled data packet size to the required data packet size for all users or data flows. Upon receiving the QoS adjustment suggestions from RAN, CN or other management entity will re-calculate the QoS for each layer and informs the QoS parameters to other layers.

4) For each user or data flow, the RAN calculates how many data need to be transmitted during a certain time window to guarantee the QoS requirements and schedules the radio resource accordingly. The buffer status and scheduled data packet size within a certain time window will be used at the TCP layer to adjust the TCP transmit packet size.

5) Based on the coordination between the CN and RAN, or even between all the layers, the QoS requirements for all the layers will be finally determined before the data transmission. Alternatively, a powerful management center in the network is responsible for joint optimization of the end-to-end QoS configuration and configures QoS for each layer. This will facilitate deterministic end-to-end QoS provision.

Note that during data transmission, the UE feeds back the end-to-end QoS status to the wireless network, which can serve as an important reference for the adjustment of the end-to-end and per-layer QoS parameters.

*To introduce QoS management in the application layer:* Traditionally the wireless QoS management is based on the requirements of the mobile services, with proper RAN resource allocation and air interface transmission schemes like channel coding, modulation, waveform, multiple access, and multiple antenna technologies. However, in the mobile communications, the mobile services also need to match the conditions of wireless networks. The application data format need to be adjusted according to the jointly determined QoS parameters at the application layer. For example, if the wireless channel does not support 1080P video for a video stream service, video packet with 720p may achieve better QoS or QoE performance. QoS flows will be generated for mobile service (e.g. FTP, video call, streaming) with different QoS requirements. This leads to more reliable service data format selection and deterministic QoS guarantee at the application layer. The transmission efficiency will be significantly improved.

*To introduce QoS management in the transport layer:* This requires the TCP has sufficient knowledge of the QoS requirements of each data flow from the application layer, and makes optimal data packet scheduling to meet the QoS requirements and the fairness metric. A context-oriented TCP design is proposed in [16], which understands the application context and adapts to varying network conditions with flow-based QoS control. This scheme introduces QoS provision before the mobile network (as traditionally implemented in the core network). However, the QoS parameters may not be accurate and more importantly may not be well supported by the mobile network. With the jointly determined QoS parameters available at the TCP layer, satisfactory QoS management will be more conveniently achieved, thus laying a solid foundation for the end-to-end QoS optimization.

*RAN information assisted deterministic TCP:* The TCP layer should be able to adjust its transmit strategy according to the wireless channel variations and scheduling results of the RAN. This mandates feedback of RAN information from the RAN to the TCP for each TCP transmit window. The information may include the scheduling results like how many data will be scheduled for each user for the next time window, the current buffer status, the channel statistics (note that it would be extremely challenging to feed back the real time channel information due to transmit latency from the RAN to TCP). Based on these RAN information, the TCP knows exactly how many data to transmit in the next transmit window to satisfy the QoS requirements. Hence, the deterministic TCP transmission is achieved. In contrast to traditional TCP schemes, transmission efficiency is significantly improved, with maximum transmit data packet size, lowest conjestion ratio, and lowest latency.

*Deterministic IP layer:* Technologies that provide determinism in the IP network have been studied in recent years, in an effort to achieve better guarantee of minimum and maximum end-to-end latency from source to destination, and bounded packet delay variation, to minimize packet loss ratios, and to minimize out-of-order packet delivery. A deterministic IP network requires a completely redesigned architecture that includes network resource layer management, deterministic IP routing layer management, and deterministic service layer management. Based on the jointly determined QoS parameters, deterministic IP technologies will further ensure QoS performance of the data packets [17].

*Prediction of the wireless channels to facilitate reliable pre-scheduling:* The TCP transmit period is much larger than the resource scheduling time interval at the RAN (usually millisecond level). This mandates reliable resource pre-scheduling in the RAN scheduler during the next TCP transmit period. Toward this end, accurate estimation and prediction of the dynamic wireless channels in temporal and frequency domains will play an essential role. Recently, artificial intelligence and deep learning technologies have been introduced in wireless channel prediction, which has exhibited large performance improvements [18]. Alternatively, resource scheduling can be implemented in transform domains, like the delay, doppler, and angular domains, where the channels' temporal correlations are relatively much higher than that in the temporal and frequency domains [19]. This enables more reliable pre-scheduling, which facilitates more deterministic and efficient end-to-end QoS guarantee.

*Network architecture to support deterministic end-to-end transmission:* Following the above design methodologies, deterministic and jointly determined QoS parameters from the application layer to the wireless communication network will be achieved. Toward this end, tight coordination or even convergence between different layers is highly motivated. The network architecture which will facilitate smooth information exchange between different layers and joint optimization will be the one important design direction. The new network architecture will also mandate significant changes to the related standardization, including new entities, functions, interface, and signaling. This will bring a paradigm shift of the mobile communications, because all the players in the end-to-end communication industry need to collaborate and the required standardization efforts will cover the traditionally different standardization bodies like 3GPP, IETF, ITU, and ISO.

A brief summery of the end-to-end design is illustrated in Fig.2.

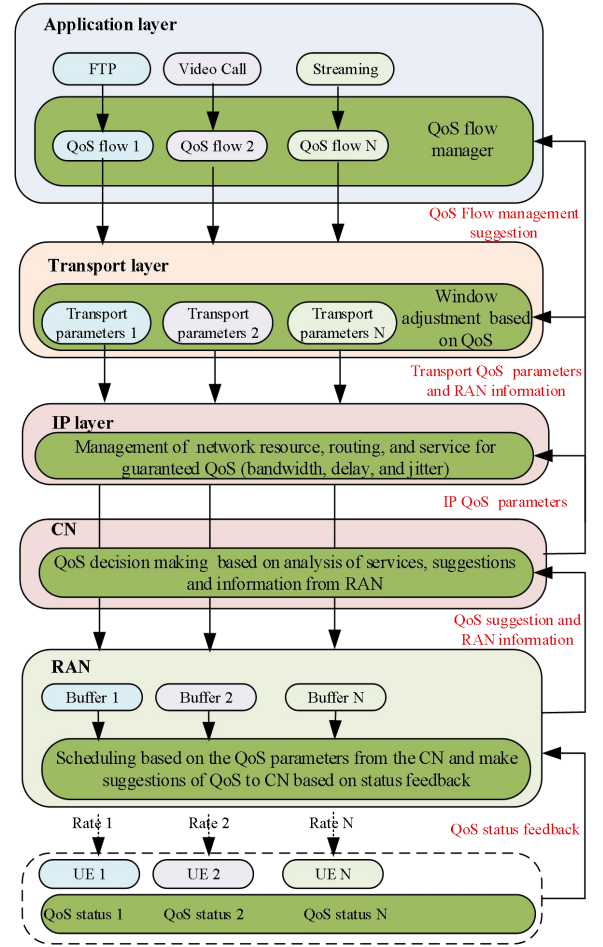

Fig. 2. End-to-end deterministic transmission for 6G and beyond

IV. CONCLUSION

In this comment article, the fundamental challenges on the end-to-end QoS guarantee in wireless communication networks like 5G are analyzed. The inherent loose coordination between different layers leads to unreliable data transmission with uncontrollable packet delay and packet error rate, with terrible waste of network resources incurred for data re-transmission. To tackle these challenges, design methodologies for deterministic end-to-end 6G and future wireless communication networks are given, which will bring a paradigm shift to converged design of different layers in the OSI service model.